\documentclass[preprint,12pt]{elsarticle}
\usepackage[utf8]{inputenc} % added by arXiv
\usepackage{amsmath,amssymb,amsthm} % For including math equations, theorems, symbols, etc

\usepackage{comment}
\usepackage{algorithm}
\usepackage{algorithmic}
\usepackage{verbatim} %for multiline comments
\usepackage[normalem]{ulem}
\usepackage{graphicx, subcaption}
\usepackage{tabularx}
\usepackage{url}

\newcommand{\eg}{{\textit{ e.g. }}}
\newcommand{\ie}{{\textit{ i.e. }}}

\newcommand{\vs}{{\textit{ vs. }}}

\usepackage{lineno}

\journal{}

\begin{document}

\begin{frontmatter}



\title{Computing the Relative Value of Spatio-Temporal Data in Wholesale and Retail Data Marketplaces}


\author[affiliation1]{Santiago Andrés Azcoitia} 
\author[affiliation2]{Marius Paraschiv}
\author[affiliation2]{Nikolaos Laoutaris}

\address[affiliation1]{IMDEA Networks Institute, Univ. Carlos III, Leganes (Madrid) Spain}
\address[affiliation2]{IMDEA Networks Institute, Leganes (Madrid) Spain}

\date{May 2020}

\begin{abstract}
Spatio-temporal information is used for driving a plethora of intelligent transportation, smart-city, and crowd-sensing applications. Since data is now considered a valuable production factor, data marketplaces have appeared to help individuals and enterprises bring it to market to satisfy the ever-growing demand. In such marketplaces, several sources may need to combine their data in order to meet the requirements of different applications. In this paper we study the problem of estimating the relative value of different spatio-temporal datasets combined in wholesale and retail marketplaces for the purpose of predicting demand in metropolitan areas. Using as case studies large datasets of taxi rides from Chicago and New York, we ask questions such as "When does it make sense for different taxi companies to combine their data?", and "How should different companies be compensated for the data that they share?". We then turn our attention to the even harder problem of establishing the relative value of the data brought to retail marketplaces by individual drivers. Overall, we show that simplistic but popular approaches for estimating the relative value of data, such as using volume, or the ``leave-one-out'' heuristic, are inaccurate. Instead, more complex notions of value from economics and game-theory, such as the Shapley value need to be employed if one wishes to capture the complex effects of mixing different datasets on the accuracy of forecasting algorithms. Applying the Shapley value to large datasets from many sources is, of course, computationally challenging. We resort to structured sampling and manage to compute accurately the importance of thousands of data sources. We show that the relative value of the data held by different taxi companies and drivers may differ substantially, and that its relative ranking may change from district to district within a metropolitan area.
\end{abstract}


\end{frontmatter}

%%\linenumbers

\section{Introduction}
\label{sect:introduction}

Data-driven decision making is bringing significant improvements to many sectors of the economy, including in several applications related to ubiquitous computing in the areas of transportation, mobility, and crowd-sensing. A solid body of research has studied matters of route optimization and city infrastructure planning~\cite{Xia19, Zhao19, Fang18, Yan18, Zhang19}, whereas companies like Uber are increasingly deploying and operating sophisticated systems for optimising their operations using live data\footnote{Examples of how Uber leverages spatio-temporal data across its main processes and operations~\url{https://eng.uber.com/forecasting-introduction/} (last accessed May 2020).}. Such models and algorithms often require combining data from different sources, and sometimes, even fusing together data from different domains~\cite{Zheng15}. 

Data is by now considered a key production factor, comparable in importance to labour, capital, and infrastructure. Companies often need data that they cannot collect on their own, and for this they resort to commercial data marketplaces. There are different types of marketplaces. Personal Information Management Systems (aka PIMS, like Digi.me, MyDex, GeoDB, HAT, EarnieApp, Citizen.me, and Meeco\footnote{See \url{https://digi.me/}, \url{https://mydex.org/}, \url{https://geodb.com/en/},  \url{https://www.hubofallthings.com/}, \url{https://ernieapp.com/}, or \url{https://www.meeco.me/}, last accessed May 2020}) allow individuals to sell their personal data, whereas general purpose (DAWEX or AWS data exchange\footnote{See \url{https://www.dawex.com/ and https://aws.amazon.com/es/data-exchange/}, last accessed May 2020}) and domain specific marketplaces for marketing (Openprise, Lotame PDX\footnote{See \url{https://www.openprisetech.com/data-orchestration/data-marketplace/} and \url{https://www.lotame.com/pdx/}, last accessed May 2020}), business intelligence (Qlik\footnote{See \url{https://www.qlik.com/es-es/products/qlik-data-market}, last accessed May 2020}), trading and investment information (Battlefin\footnote{See \url{https://www.battlefin.com/}, last accessed May 2020}) allow companies to sell to other companies in a B2B manner real-time\footnote{See \url{https://streamr.network/}, \url{https://data.iota.org/}, \url{https://airbloc.org/}, and \url{https://geodb.com/en/}, last accessed May 2020}, or siloed offline datasets.  

In almost all commercial marketplaces, pricing is left to sellers and buyers to agree. Sellers may set a fixed price, or let buyers bid for data~\cite{Mehta19}, or even do a combination of the two. Such empirical pricing operates with minimal information, namely a high level description of the dataset, including the number of data points it includes. Recent research efforts have attempted to evolve data pricing towards a more principled basis. For example~\cite{Agarwal19, Chen19, Paraschiv19} attempt to associate the value of data with the context and the task for which it will be used. Overall, the buyer is better positioned to estimate the actual monetary worth of data because he can calculate how an improved accuracy in a revenue generating task, such as product- or content-recommendation, translates into actual money. For example, eBay estimated that a 15\% improvement on the recommender system translated into 6\% increase in revenues (\ie an increment of \$0.54 billion in 2016)~\cite{Brovman16}. A bid can be placed for an individual dataset, or for a combined one that uses multiple sources to, e.g., increase coverage in time, space, or some other dimension. In this latter case, it is an open problem how to split an accepted bid among the different contributors to the combined dataset.  

\section{Our Contributions}
\label{sect:Contributions}

Our work looks at this open problem for the case of spatio-temporal data. In particular, we study how to compute the relative value of different spatio-temporal datasets used in forecasting future demand for a service across space and time in a metropolitan area. Companies already offering service in overlapping areas can, for example, pool together their data to increase the accuracy of forecasting and its coverage. Improved forecasting can be used by the same companies to improve operations, such as dispatching vehicles, or provisioning service points. It can also be sold to successful bidders. In the latter case, the relative value that we compute for each contributing source provides a fair way for splitting accepted bids among them.  

For the purpose of our work, we concentrate on vehicle-for-hire demand prediction in Chicago and New York. While our examples and findings are specific to this particular urban mobility use case, the methods that we develop for assigning value to spatio-temporal datasets held by (taxi) companies and individuals (drivers) are more general in scope, and can thus be used in other use cases beyond transportation, such as tourism, health services, entertainment, energy or telecommunications. We will assume that drivers sell their data via retail data marketplaces, like the various PIMS mentioned earlier, whereas taxi companies make them available via wholesale marketplaces for B2B data.  

We develop data valuation methods to answer a series of fundamental questions pertaining to both wholesale and retail data fusion. For example, ``Does combining multiple datasets of past taxi rides always benefit the forecasting accuracy of future services?''. Also, when it does, ``How should we attribute the improved forecasting precision to the individual datasets used to produce it?''.

To answer the above questions, we use the Shapley value~\cite{Winter02} from collaborative game theory as a baseline metric for establishing the importance of individual \textit{players} (be they taxi companies or individual drivers) in the context of a \textit{coalition} of data providers. The Shapley value has many salient fairness properties and wide market adoption, but at the same time entails serious combinatorial complexity challenges since its direct computation in a coalition of size $N$ requires enumerating and calculating the value of $O(2^N)$ sub-coalitions. This may be possible for a few tens of data providers, which is the case of companies in wholesale markets, but becomes impossible when considering hundreds or thousands of them in a retail data market setting.

Furthermore, we look at the tradeoff between fairness and scalability / practicality by studying and comparing against simpler heuristics used to estimate the value of data, based on:

\begin{itemize}
    \item \textit{data volume}, in our case, taxi rides. This has been used in marketplaces trading  marketing or user profiling data~\cite{Mehta19}. While certainly more practical, the latter assumes that any reported ride from the past has equal value for predicting rides of the future. 
    \item \textit{leave-one-out (LOO)}. LOO has been used for ``denoising'' datasets, by omitting data points that reduce the accuracy of a machine learning algorithm~\cite{Ghorbani19}. Unlike the Shapley value that requires enumerating $O(2^N)$ sub-coalitions, LOO is examining only a single sub-coalition per source.
\end{itemize}

\noindent \textbf{Findings:}
We first study data fusion at the granularity of entire companies. Since the number of such companies covering the same geographical area is typically small, the relative value of their data can be computed directly from the definition of what the Shapley value is. This, however, becomes infeasible at the level of individual taxi drivers, since the latter may amount to several thousands for large metropolitan areas. To address this issue, we compare different approximation techniques, and conclude that structured sampling~\cite{Fatima08} performs much better than other approaches such as Monte Carlo~\cite{Stanojevic10, Ghorbani19} and random sampling. 

By applying our model and valuation algorithms to taxi-ride data from Chicago and New York, we find that sufficiently large companies hold enough information to independently predict the overall demand, at city-level, or in large districts, with over 96\% accuracy. This effectively means that inter-company collaboration does not make much sense in such cases. On the other hand, when the objective is to make predictions at a finer -- district-level -- granularity, then there are plenty of districts in which companies have to combine their data in order to achieve a sufficient forecasting accuracy. We compute the relative value of different contributions in such cases by computing the Shapley value for each taxi company.  We find that there exist companies whose values differ by several orders of magnitude, and that the importance of the data of a given company can vary as much as $\times 10$ across districts. More interestingly, the Shapley value of a company's dataset does not correlate with its volume, \ie there are companies that report relatively few rides but have a larger impact on the forecasting accuracy than companies that report many more rides. The LOO heuristic also fails to approximate the per company value as given by Shapley.

Similar phenomena are observed at the finer level of individual drivers. We show that by combining data from relatively few drivers one can easily detect peak hours at city level. At district level, however, more data needs to be combined, and this requires making use of our fastest approximations for the Shapley value based on structured sampling. \emph{Overall our work shows that computing, even approximately, the Shapley value is a ``necessary evil'' if one wants to split fairly the value of a combined spatio-temporal dataset.}

\section{Background}
\label{background}

\subsection{Definitions and problem statement}
\label{sect:problem description}

Let $N$ denote a set of data sources, each one contributing a dataset $S_n, n\in N$. A dataset is a set of spatio-temporal observations $(x,t)$ denoting the spatial ($x$) and temporal ($t$) coordinates where demand for a service has taken place. All sources report spatio-temporal demand observations over a common period. The dataset is then split, along the time dimension, into an \emph{observation period}, $T_{\text{o}}$ and a \emph{control period}, $T_{\text{c}}$.   

The segment of the original dataset, corresponding to $T_{\text{o}}$, constitutes the training set, while the one spanning $T_{\text{c}}$, represents the test set. Throughout the paper, we train a predictive algorithm on subsets of the complete training set (containing a part of the total number of data sources) and perform predictions on the test set. The accuracy of the trained model is gauged by a series of similarity metrics, applied to the predictions and test set ground truth, respectively. The said similarity metrics allow one to define the notion of \emph{value} of a dataset $S_K$, where $K \subseteq N$, which we denote as $v(S_K)$. Thus, $v(S_K)$ represents the accuracy of the predictive model, according to the chosen metric, when training is performed on the data from all sources $k \in K$, and prediction is performed on the fixed test set, containing data from all sources.

Our objective is to find a value assignment method, $\mu(n_i)$, that captures the relative importance of the data originating from source $n_i$ to the predictions of the forecasting algorithm. The value assignment method $\mu(n_i)$ would thus be a function of the dataset of source $n_i$, the datasets of all other sources other than $n_i$, and v.

\begin{equation}
\mu(n_i) = f(S_{n_i}, S_j, v), j \in N - \{n_i\}
\label{main_val_func}
\end{equation}

\subsection{Spatio-temporal forecasting}
\label{sect:Model}

In our definition, the notion of \emph{data value} is directly linked to the similarity metric one uses in order to compute the model accuracy. As such, we ensure the robustness of our results by considering three appropriate metrics, namely: cosine similarity, numerical similarity and dynamic time warp. 

Let $\hat{S}_K[t]$, with $t \in T_{\text{c}}$ and $K \subseteq N$, be the model prediction on the control period, after being trained on $S_K[t]$, with $t \in T_{\text{o}}$, and $S_N[t]$, with $t \in T_{\text{c}}$, be the test set. The value of the data produced by the source subset $K$, can be defined in terms of the \emph{Cosine Similarity} as:

\begin{equation}
v_1(S_K) = \text{CosSim}(S_N[t], \hat{S}_K[t]) = \frac{\sum_{t \in T_\text{c}}{S_N[t] \cdot \hat{S}_K[t]}}{\sqrt{\sum_{t \in T_\text{c}}{(S_N[t])^2}} \cdot \sqrt{\sum_{t \in T_\text{c}}{(\hat{S}_K[t])^2}}}
\label{eq:CosSim}
\end{equation} 

\begin{figure}[tb]
  \centering
  \includegraphics[width=300pt]{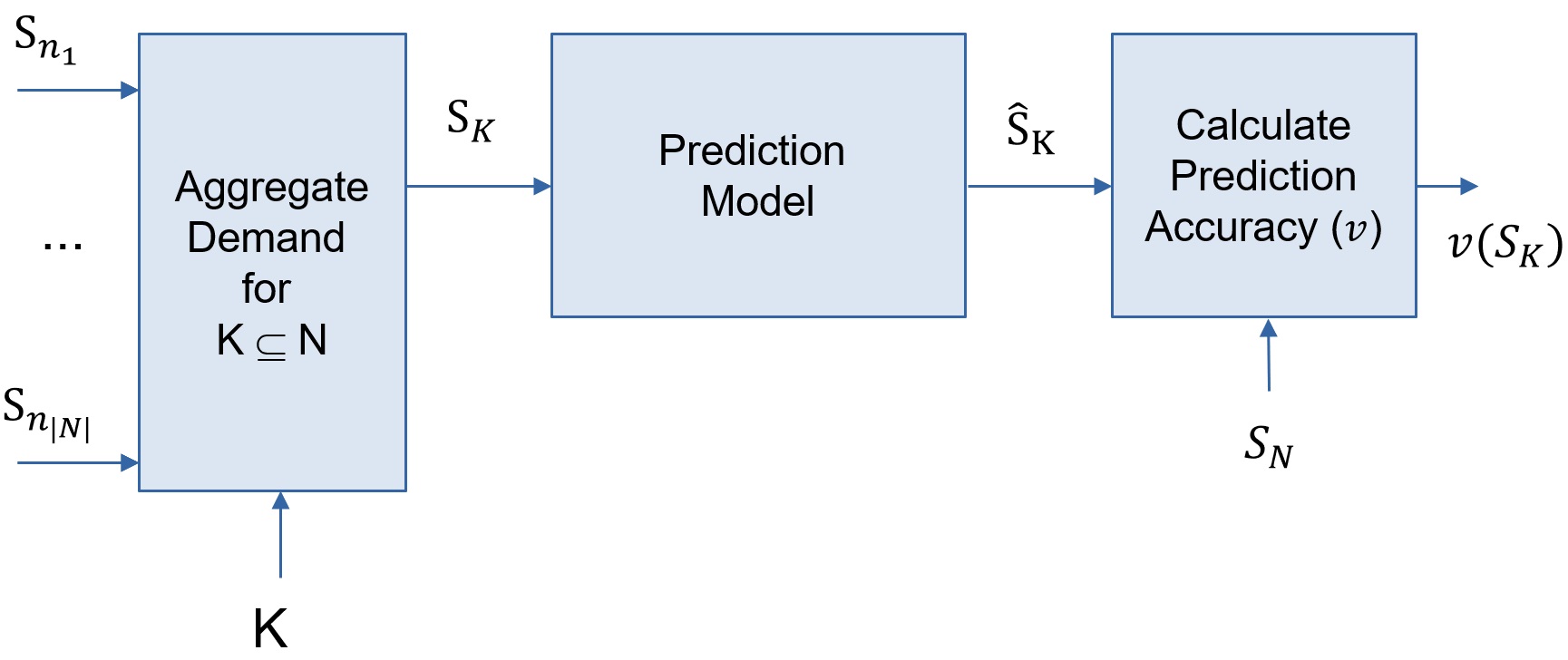}
  \caption{Block diagram describing the general prediction model used throughout the paper. The model constructs the aggregate input (or training) demand $S_K$ from a set of sources $K \subseteq N$ over an observation period $T_o$. This input then drives a prediction model, which in our case is a multi-seasonal SARIMA algorithm with hourly, daily and weekly sub-components. As a result, the prediction model produces a forecast $\hat{S}_K$  over the control period $T_c$ which is compared to the ground truth, as described in Eq.~\ref{eq:CosSim}.}
\label{Figure0}
\end{figure}

Figure \ref{Figure0} shows a block diagram that describes the general prediction model used throughout the paper. The model can easily accommodate other similarity measures as the value function. 

The cosine similarity has been chosen, among other similarity metrics, due to the fact that it preserves the direction of vectors irrespective of their overall length. As we are dealing with relatively long vectors, element-wise comparison metrics, such as Euclidean or Hamming distances, would not be appropriate, as even a slight shift could cause the error estimate to increase drastically. Other metrics, such as Mean Squared Error and Root Mean Squared Error have been rejected due to the fact that these tend to artificially favor large errors and mask smaller ones. The second considered distance metric is \emph{Numerical Similarity}, defined as:

\begin{equation}
    v_2(S_K) = 1 - \frac{1}{n} \cdot \sum_{t \in T_c}{\frac{|\overline{S_N}[t]-\overline{\hat{S}_K}[t]|}{\overline{S_N}[t]+\overline{\hat{S}_K}[t]}},
\label{eq:NumSim}
\end{equation}
which works with normalized output ($\overline{S_K}$) and is thus module-independent. The averaging performed does help in reducing inconsistencies introduced by the element-wise comparison. The third metric chosen is \emph{Relative Dynamic Time Warp} or RDTW, defined as:

\begin{equation}
    v_3(S_K) =  1 - \frac{\text{DTW}(\overline{S_N}[t], \overline{\hat{S}_K}[t])}{\text{DTW}(\overline{S_N}[t], 0)}
    \hspace{1cm}
    t \in T_c,
\label{eq:RDTW}
\end{equation}
a similarity metric often used in signal processing and automatic speech recognition, where robust comparison of time signals is required.

\subsection{Introducing the Shapley value}
\label{sect:Introducing Shapley}

Establishing individual player contributions to a collaborative game has long been a central problem of cooperative game theory. To this end, Shapley proposed that a player's value should be proportional to their average marginal contribution to any coalition they may join \cite{Winter02}. In what follows, we adapt the \emph{Shapley value} definition to our use case.

Let $N$ be a set of sources and $S_N$ be their aggregate data, with a value $v(S_N)$. The \emph{Shapley value} is a uniquely determined vector of the form $(\phi(n_1),..., \phi(n_{|N|}))$, $n_1 ... n_{|N|} \in N$, where the element representing source $n_i$ is given by

\begin{equation}
\phi(n_i) = \sum_{K  \subseteq {N \setminus \{n_i\}}} \frac{|K|!(|N| - |K| - 1)!}{|N|!} [v(S_{K} \cup S_{n_i}) - v(S_K)],
\label{shapley_eq}
\end{equation}
where $K \subseteq N \setminus n_i$ takes the value of all possible coalitions of sources, excluding $n_i$. $v(S_{K} \cup S_{n_i})$ represents the value of the combined data from the $K$ sources and source $n_i$.

We can use the Shapley value, according to Eq.~\ref{main_val_func}, as a credit assignment method, where $\mu(n_i) = \phi(n_i)$. To determine the individual values $v(S)$ and present the results of our experiments, we have employed cosine similarity as our metric of choice, and we have validated the obtained outcomes against the other two metrics, for which we have obtained agreeing results (see appendix \ref{sect:validatingOthermetrics}).

\subsubsection{A toy example}

Consider a group of taxi companies agreeing to pool together their spatio-temporal data, containing demand for taxi rides within a city. One method to determine the value of a company is to observe how well the company is able to reconstruct the total aggregate, that is, the data coming from all companies, by solely using its own. As such, the data of one single company, or a group thereof, is used to train a predictive model, and the reconstruction error, between the prediction and an actual ground truth, is measured. This error, or rather its opposite, the reconstruction accuracy, represents the value of the company (or coalition of companies). 
Aggregation leads to a highly non-trivial behavior of the value function, and in the following, we will discuss a few particular cases in more detail. 

\begin{figure}[tb]
\begin{subfigure}{.3\textwidth}
\centering
\includegraphics[width=5cm,height=4.3cm]{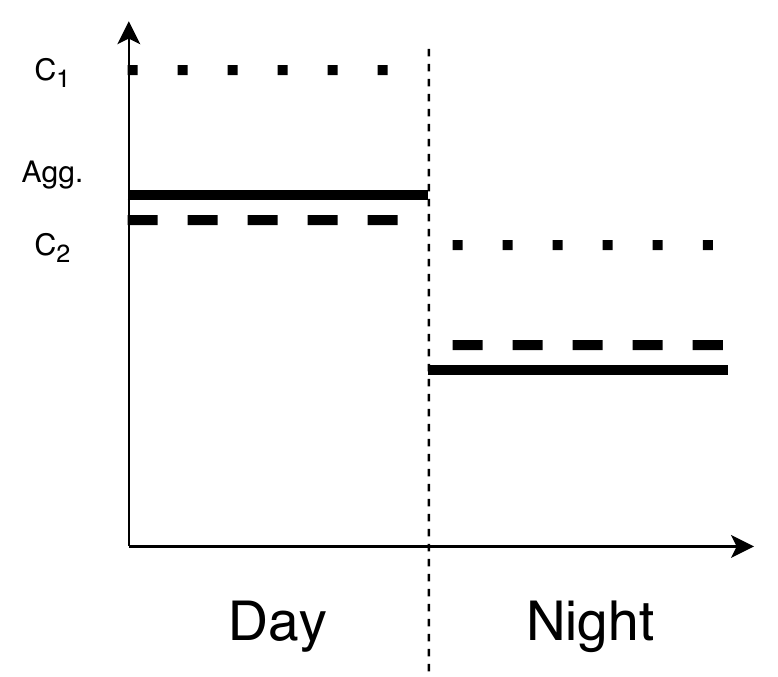}
\caption{}
\label{fig_a}
\end{subfigure}\hfill
\begin{subfigure}{.3\textwidth}
\centering
\includegraphics[width=5cm,height=4.3cm]{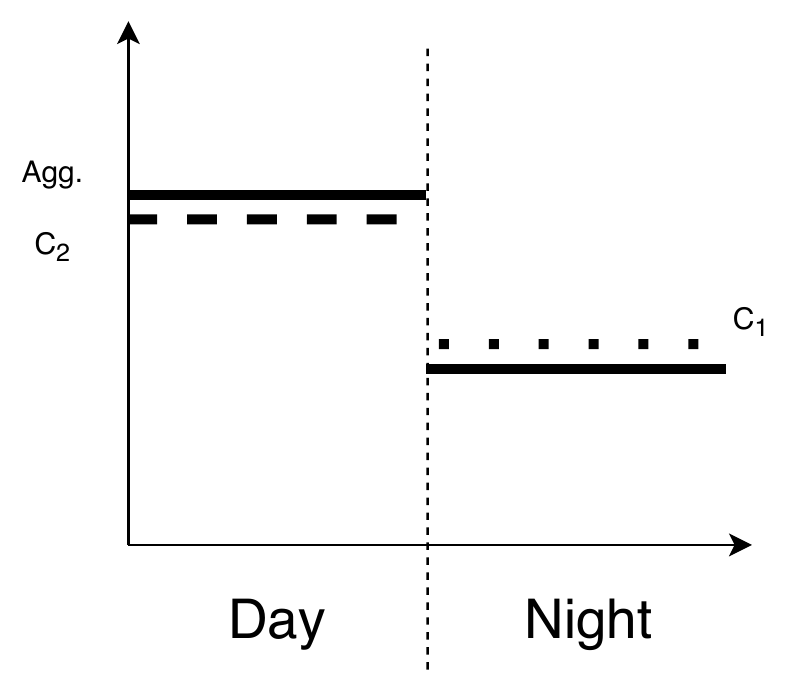}
\caption{}
\label{fig_b}
\end{subfigure}\hfill
\begin{subfigure}{.3\textwidth}
\centering
\includegraphics[width=5cm,height=4.3cm]{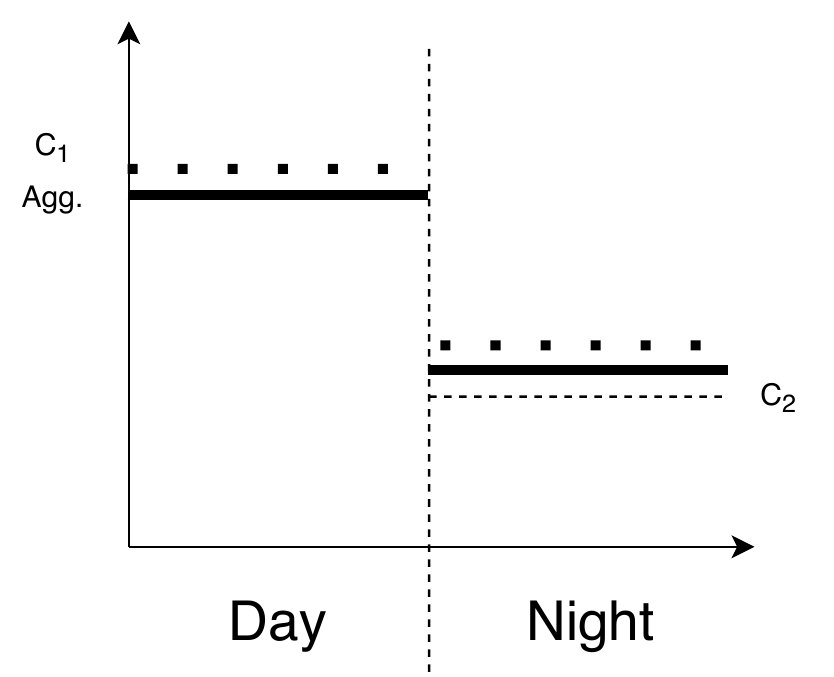}
\caption{}
\label{fig_c}
\end{subfigure}\hfill
\caption{Data aggregation can influence its value in nontrivial ways. In figure (a) we have two companies, $C_1$ and $C_2$, active during an entire day. As $C_2$ has an average value closer to the reference aggregate (which represents data coming from many other companies), its data is better able to reconstruct the aggregate than $C_1$'s data. In figure (b), the two companies act at different hours, and neither will be able to reconstruct the complete output by itself, but by combining their data, they gain a significant advantage. In figure (c), the combination of data from a company active during the entire day with that of one only active during the night, will be detrimental to the task of predicting the complete aggregate, as $C_2$'s data will distort the true activity gap between day and night. Not all data aggregations lead to added value.}
\end{figure}

Let us consider a toy example, depicted in Fig.~\ref{fig_a}. A  number of companies combine their data, to produce a spatio-temporal output or signal (continuous line), representing the total aggregate demand. For simplicity, the time scale is that of a single day, split into day-time and night-time, and also all signals are drawn as constant. Companies whose overall behavior is closer to the \textit{average} may be able to predict the complete aggregate signal by themselves, without a need to form coalitions with other companies. As such, their value will be ranked high, by our algorithm. In the example, company $C_1$ is less valuable than $C_2$, as the signal of $C_2$ better emulates the total aggregate.

In the same setting, we also discuss the problem of complementarity, depicted in Fig.~\ref{fig_b}. Company $C_1$ is only offering its transport services during the night, while company $C_2$ is active solely during the day. Taken individually, the data of neither of these two is able to reconstruct the complete aggregate, however, they gain tremendous value as a coalition. Indeed, by combining their data, the resulting signal covers the entire time-span of the aggregate.

Data aggregation, however, does not always lead to an increase in value. Indeed, there are cases where adding a new provider to a coalition reduces the total value. A simple example is presented in Fig.~\ref{fig_c}, one company, $C_1$ provides data spanning the entire day (both day-time and night-time), and is also close to the total aggregate, while the other, $C_2$, only provides data during the night. The predictive accuracy of both data sets combined is lower than that of $C_1$,  because the absence of reports for day from $C_2$ will make most estimators believe that the traffic intensity gap between day and night is smaller than the real one.

It is thus clear that, depending on the particular characteristics of different datasets, mixing data may or may not be beneficial. 

\subsubsection{Computing the Shapley value}
\label{sect:computing Shapley of individuals}
Unfortunately, the Shapley value has also been proven to be NP-hard for many domains \cite{Bachrach09}. Since it takes into account all possible coalitions, for each source, the number of terms scales with $2^{|N|}$, where $|N|$ represents the number of sources, therefore it quickly becomes computationally unfeasible.

In Ref.~\cite{Stanojevic10} the authors use Monte Carlo to approximate the Shapley value for computing the cost contribution of individual households to the peak hour traffic and costs of an Internet Service Provider (ISP). Other recent works have presented approximation algorithms for Shapley for specific problems of lower complexity \cite{Cabello18, Zhao18}. Here we have tested both Monte Carlo and Truncated Monte Carlo methods, as well as Random Sampling and various Structured Sampling techniques (see appendix \ref{sect:EvaluatingApproxAlgorithms} for more information):

\begin{enumerate}
\item Truncated Monte Carlo approximation (TMC)
\item Random sampling (RS)
\item Structured sampling (SS), which plans the sampling upfront to ensure that all players appear $r$ times in each position of the $r \cdot |N|$ sampled permutations of N.
\item Truncated SS (TSS), which is a variant of SS that stops computing sample permutations once the accuracy reaches a certain threshold (\eg 95\% of $v(S_N)$) 
\end{enumerate}

Having evaluated the above algorithms extensively (see details in appendix \ref{sect:ShapleyApproxAlgorithms}) in terms of precision and robustness vs. computational time, we have selected the TSS algorithm since it achieved the best trade-off on all the datasets we used for testing.

\subsection{Simpler heuristics for value estimation}

One might initially think that the value of the data coming from a provider $n_i$  is given by its volume. In fact, some data marketplaces, which trade marketing and user profiles, establish the price of their datasets proportionally to their volume. We will also consider value distribution based on data volume, which results in the value assignment metric $\mu(n_i) = |S_{n_i}|$, where $|S_{n_i}|$ stands for the data volume of source $n_i$, or the number of data points originating from this particular source.

The \emph{Leave One Out} (LOO) method, widely used in various areas of machine learning, considers that the value of a source $n_i$ is the difference in performance when the data corresponding to that particular source is removed from the training set. We define the LOO value of source $n_i$ as $LOO(n_i) = v(S_N) - v(S_{N-\{n_i\}})$. In accordance with Eq.~\ref{main_val_func}, the value assignment method in this case, is provided by $\mu(n_i) = LOO(n_i)$.

LOO can be computed in $O(|N|)$ time and has proven to be valuable for optimizing the outcome of an algorithm by trimming data with negative $LOO$ values \cite{Ghorbani19}. We will address the question of how appropriate LOO is for value assignment in the next sections.

\section{Computing the importance of data in wholesale collaborations}
\label {sect:Wholesale}

We start with the case that different companies pool together their data to improve the quality of their demand forecasts, either for their own use, or to sell it to an external buyer that has offered an accepted bid. In both cases, it is relevant to know how important the data contribution of each taxi company is.

\subsection{Description of the setting and assumptions}
\label{sect:wholesale setting}
For the purpose of this use case, we will focus on metropolitan vehicle-for-hire markets and we will assume that i) service demand observations will be taxi rides reported in a certain spatial coordinates at a certain time, and ii) data sources will be the databases of taxi companies that contain a log of such taxi rides. Our objective will be to forecast the aggregated demand in a control period taking as an input the demand reported in an observation period. Increasing the accuracy of such a prediction model is important both for operational needs (\eg knowing where to dispatch drivers in anticipation of demand) and planning issues (\eg deciding where to place taxi service points), so as it would make sense for a company to collaborate in case its prediction accuracy could be significantly improved by pooling similar data with other companies.

In order to compute results for a real scenario, we will make use of a public dataset of taxi rides from the city of Chicago, \footnote{see \url{https://data.cityofchicago.org/Transportation/Taxi-Trips/wrvz-psew}, last accessed May 2020)} which is a log of taxi rides that licensed companies report to local regulatory bodies. This dataset consists of more than 94 million rides from 160 companies, spanning from 2013 to 2019. We will filter data for the first half of 2019 for the analysis (see Table \ref{tab:table0} for a summary of the properties of this dataset). We will consider the demand for the main 15 taxi companies in that city, plus an additional hypothetical 16th company, where we aggregate the information from the rest of companies, which account for less than 5\% of the total demand. In section \ref{sect:TaxiNYC} we will also present the results for a similar data set from New York City.

\begin{table}
  \caption{Chicago City taxi rides dataset (retrieved during Nov'19). Brief description and statistics for the whole data set and for the specific period which was used in the simulations.}
  \begin{tabular}{lp{5cm}p{5cm}}
    \hline
    Time period & 01-01-2013 - 09-01-2019 & 01-01-2019 - 09-01-2019 \\
    \hline
    Rides&94 millions&11.1 millions\\
    Companies&160 with 101 individual licenses&58. 94\% rides from top 15 companies\\
    Districts&\multicolumn{2}{c}{77 districts (administrative communities) of Chicago City}\\
    Taxi Ids&19,014. 55\% of the total licenses associated to 5 companies&6,469\\
    \hline
\end{tabular}
\label{tab:table0}
\end{table}

We will start our analysis by first checking the cases that make collaboration between companies meaningful. For those cases, we will then compute a fair measure of the importance of each individual company based on the quality of the data it offers. We will look at those two matters at both city level, as well as independently for each of the 77 different administrative areas (hereinafter, districts) in which Chicago is divided.

\subsection{Demand forecasting at city level}
\label{sect:wholesale city}

\begin{figure}[tb]
  \centering
  \includegraphics[width=\linewidth]{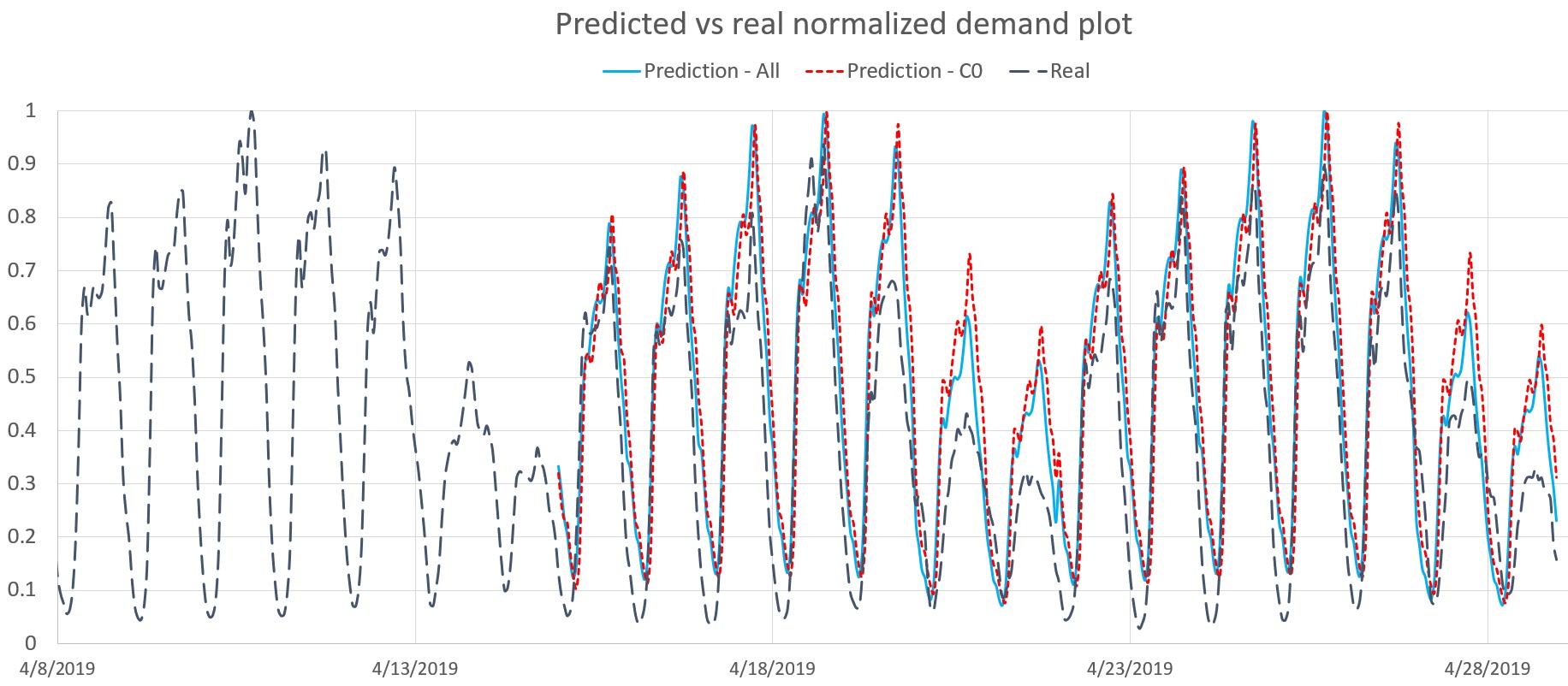}
  \caption{Example plot of a city-wide SARIMA model fit using the information from all companies and only from company labelled as C0 }
\label{Figure1}
\end{figure}

Figure \ref{Figure1} shows a prediction sample for a control period between Apr 15th and Apr 28th 2019 based on the observations of the previous weeks. It compares the real observed demand to the predicted demand using information from \textit{all} companies and only from the company labelled as \textit{C0}. Similar plots are obtained for the rest of the companies.
Table \ref{tab:table1} shows a summary of the forecast accuracy achieved by using all the information available and by using only the information from each company. According to our results, the demand prediction that each company is able to produce on its own yields, in general, an accuracy above 96\% at city level. This means that all companies have enough data to independently predict the future demand with at most a 4\% maximum average error. Granted that all companies have sufficient data to perform demand prediction accurately on their own, the incentives for collaboration via pooling their data together are very small. 

\begin{table}
  \caption{Accuracy metrics for example city-wide SARIMA model fit}
  \centering
  \begin{tabular}{lclc}
    \hline
    Co&Accuracy&Co&Accuracy \\
    \hline
    All&0.9833&C8&0.9797\\
    C0&0.9686&C9&0.9861\\
    C1&0.9835&C10&0.9829\\
    C2&0.9794&C11&0.9659\\
    C3&0.9737&C12&0.9845\\
    C4&0.9801&C13&0.9725\\
    C5&0.9736&C14&0.9767\\
    C6&0.9800&C15&0.9724\\
    C7&0.9804\\
  \hline
\end{tabular}
\label{tab:table1}
\end{table}

\subsection{Demand forecasting at district level}
We performed a similar analysis by isolating the rides of each of the 77 districts of Chicago. Estimating the future demand in this case becomes more challenging and, as we will show soon, often requires collaboration between different companies. 

\begin{figure}[tbp]
  \centering
  \includegraphics[width=\linewidth]{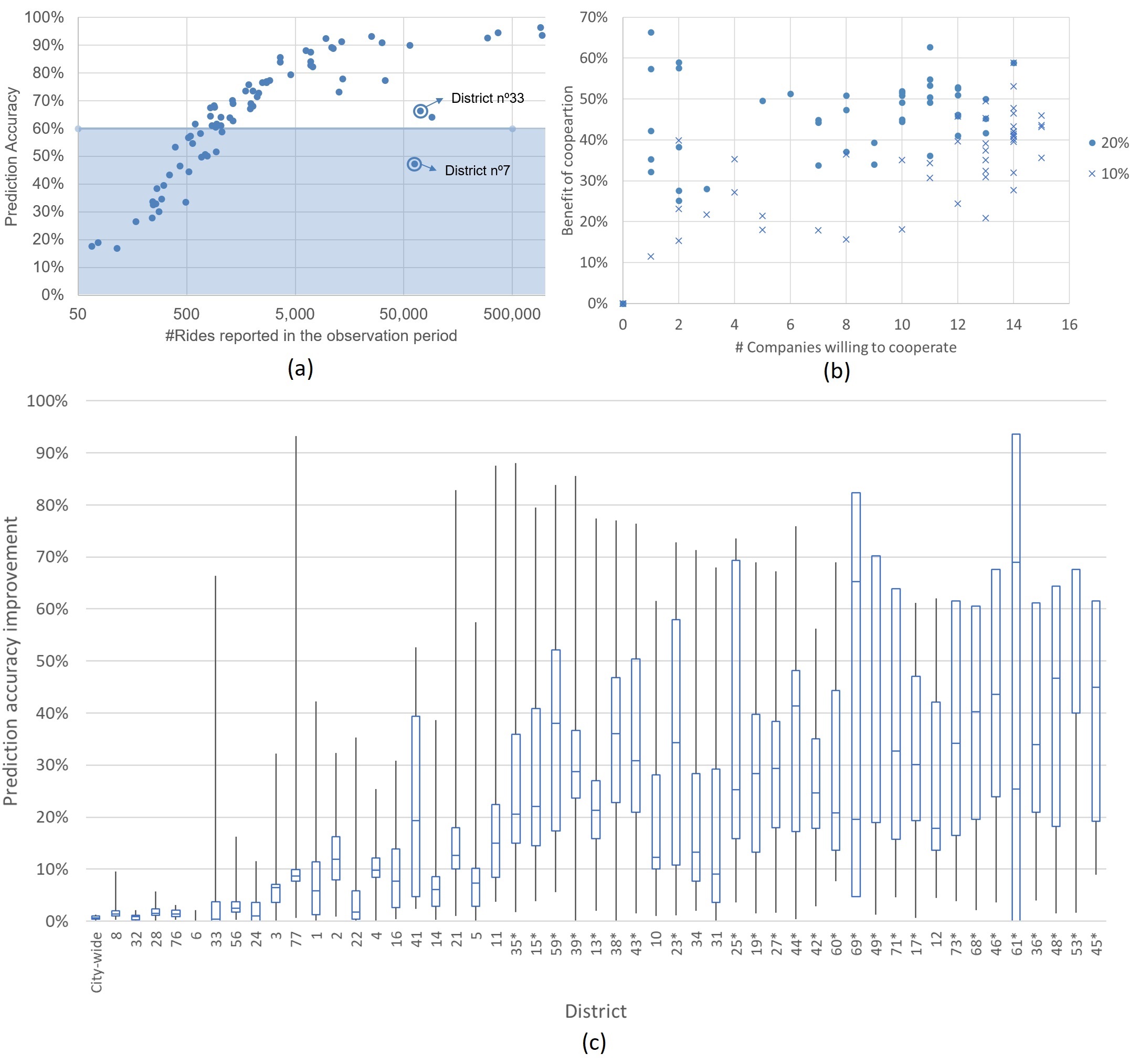}
  \caption{Demand prediction at district level. (a) Relationship between the level of accuracy achieved by district and the number of rides reported. Each point in the plot represents a district. (b) Benefit of cooperation \vs number of companies willing to cooperate in obtaining a better prediction for two different \textit{cooperation thresholds} (20\% or 10\%) (c) Potential prediction accuracy improvement by cooperation at district level. }
\label{Figure2}
\end{figure}

Figure \ref{Figure2} (a) shows the relationship between the forecast accuracy and the number of rides reported within a district. Not surprisingly, we see that the accuracy is higher in districts with a higher number of reported rides. District-level predictions are more susceptible to irregular local events than city-wide predictions. For instance, despite being one of the districts with the highest number of reported rides, district number 7 (Lincoln Park), appears to be an outlier in terms of accuracy in Fig.~\ref{Figure2} (a). While analysing manually the dataset we found out that a large number of the reported rides were due to a one time event -- a James Bay concert at the Riviera Theater, on March 19th. The resulting irregular spike that evening largely explains why the forecasting accuracy remains lower than other districts with smaller volume of demand but more regular patterns.

Another interesting case is district 33 (Near South Side), where the NFL Stadium, McCormick Place and different Museums and city attractions are located. Even though it is reporting a reasonably high number of rides (70k, ranked the fifth district in the city in terms of number of rides), the forecasting algorithm is unable to produce a prediction of high accuracy (goes up to 66\% accuracy even with all the available information used). This is due to the event-driven nature of demand in this area, which is not captured by the assumed SARIMA algorithm.\footnote{Areas like this may be amenable to a better prediction accuracy by more complex forecasting algorithms using contextual information but this goes outside the scope of this paper since our focusing is on judging the importance of different datasets for a (reasonable) predictor as opposed to designing the best predictor possible.}

Out of the 77 districts, the forecasting algorithm is able to achieve an accuracy above 60\% for 50 of them (those above the shaded region in subplot (a) of Fig.~\ref{Figure2}). This means that even by aggregating all the information available, the particular forecasting algorithm would not be able to predict the future demand with sufficient accuracy for 27 districts.

In order to check whether our findings at city level still hold at district level, we execute the forecasting algorithm in each of these 50 districts for all 16 companies. We compute for each one the \textit{benefit of cooperation} as the difference between the accuracy of demand forecasting using all companies and the average (across companies) forecasting accuracy achieved by each of them on their own. For our analysis, we will assume that a company would be willing to cooperate if its forecasting accuracy is improved by at least a minimum \textit{cooperation threshold}. Figure \ref{Figure2} (b) plots for each of those 50 districts the average benefit of cooperation (Y-axis) vs. the number of companies willing to cooperate (X-axis), considering two different \textit{cooperation thresholds}: 10\% and 20\%. 

Looking deeper within district data, we find that in all the districts of Fig.~\ref{Figure2} (b) there is always at least one company that is able to build a forecast model on its own which is very close to the one built by using all the data available. It is not necessarily always the same company across all districts, neither always the biggest one. Also we see that in general, smaller companies tend to benefit more from the cooperation. It is also worth noticing that lowering the \textit{cooperation threshold} leads to more companies willing to cooperate and a lower average \textit{benefit of cooperation}.

Having taken a first look at the benefits for different companies in different districts, we turn our attention to those districts where inter-company collaboration makes more sense. Figure \ref{Figure2} (c) depicts box-plots (over companies) of the forecasting accuracy improvement from collaboration (Y-axis) in each district (X-axis). Districts are sorted in descending order with respect to the total number of reported rides. We also include city-wide results at the leftmost point of the plot. The plot shows clearly that in the most popular districts (meaning that they report a large number of rides) the per-company benefits from collaboration are rather small as with city-wide results. However, as we move to smaller districts, the benefits of collaboration start increasing. It is at such areas where it makes sense for different taxi companies to pool their data together in order to achieve a higher demand forecast accuracy.  

In summary, the plot shows that:
\begin{itemize}
\item there are 17 districts where most companies are able to provide accurate demand forecasts (i.e., above 80\% of the accuracy achieved using all the information) and, consequently where there are weak incentives for companies to cooperate.
\item In 26 districts (marked with an asterisk in box plot (c) of Fig.~\ref{Figure2}) the average benefit of cooperation is at least 20\%.
\item In 33 districts the average benefit of cooperation is at least 10\%
\end{itemize}

Focusing on the districts where collaboration makes most sense, we will now show how to compute the relative importance of the data that each company brings. We will do that via the notion of the Shapley value that we introduced in \ref{sect:Introducing Shapley}.

\subsection{Computing the relative value of information at district level}
\label{sect:comparing metrics district}
For the 26 districts marked with an asterisk in Fig.~\ref{Figure2} (c), taxi companies would benefit from an increase in forecasting accuracy by combining their data. For each one of them we have computed the Shapley value of the 16 companies. To do that we used the value function from Eq.~\ref{eq:CosSim}, where the test dataset is now obtained by combining the taxi ride data of all companies active in that particular district, and the prediction is the output of the SARIMA algorithm, once trained on the taxi ride data from a particular coalition. For establishing value, based on all such coalitions, the Shapley formula from Eq.~\ref{shapley_eq} is used.

\begin{figure}[tbp]
  \centering
  \includegraphics[width=300pt]{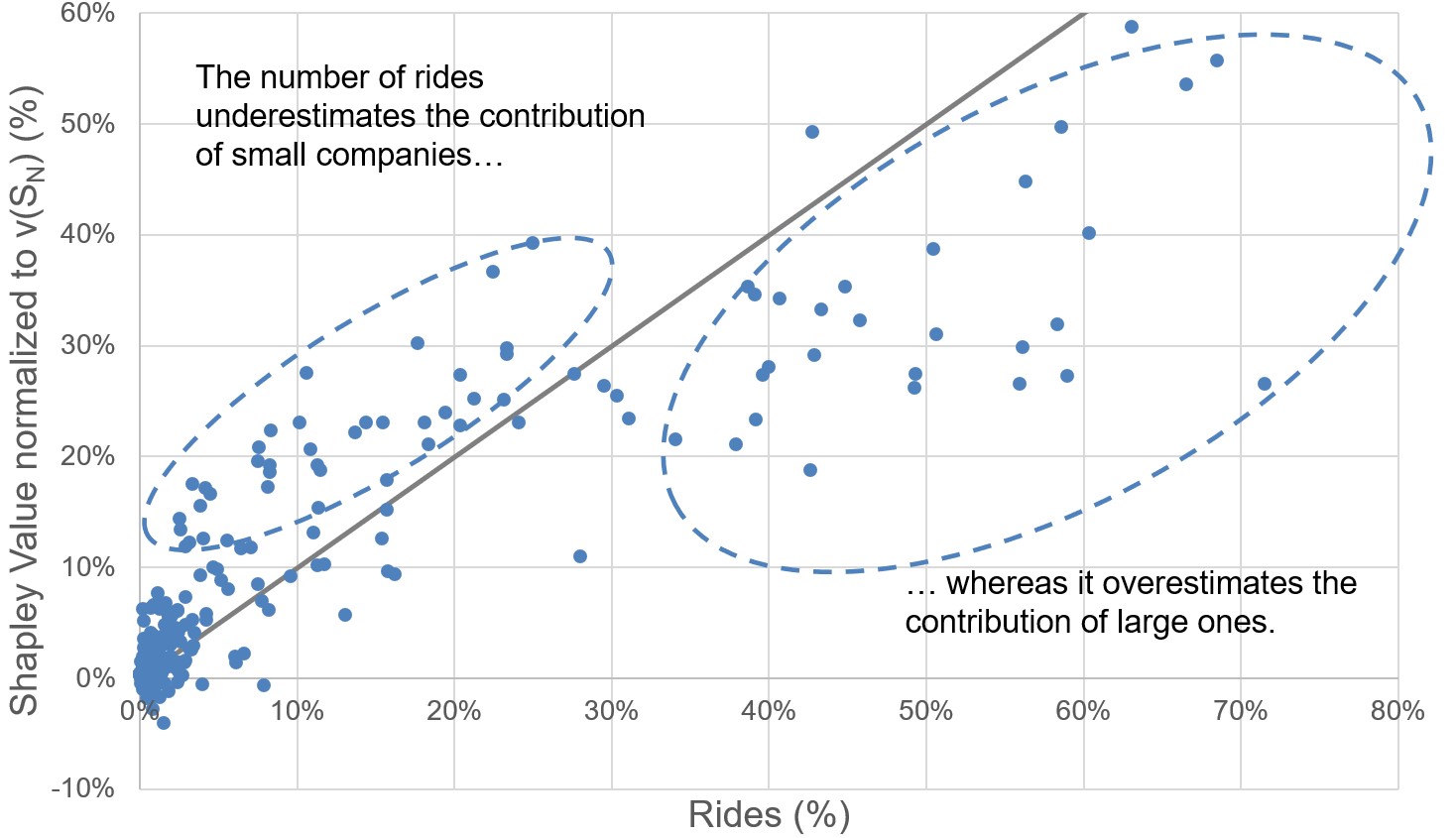}
  \caption{Shapley value vs. percentage of rides reported by company for a sample of districts. Each point represents a company in a district}
\label{Figure 5}
\end{figure}

Table \ref{tab:tableDistrictResults} summarizes the Shapley value, the LOO value and the percentage of rides reported by each company in the first 4 districts. Figure \ref{Figure 5} shows the relationship between the number of rides and the Shapley value for our forecast at district level. Each point in the plot represents a company in one of the 26 districts. The Shapley value of a company in a district, represents the average marginal contribution of its data to the obtained forecast accuracy for that district.

\begin{table}
  \caption{Shapley value, leave-one-out (LOO) value and number of rides (\%) for a sample of districts}
  \tiny
  \begin{tabular}{l||ccc|ccc|ccc|ccc}
    \hline
    &\multicolumn{3}{c}{13}&\multicolumn{3}{c}{15}&\multicolumn{3}{c}{17}&\multicolumn{3}{c}{19}\\
    Co&SV&LOO&Rides(\%)&SV&LOO&Rides(\%)&SV&LOO&Rides(\%)&SV&LOO&Rides(\%)\\
    \hline
    1&11.8&0.4&11.3&11.2&0.5&2.5&14.0&0.6&8.3&2.0&0.0&3.4\\
    2&1.8&0.2&1.2&1.8&-0.1&0.8&0.0&-0.1&0.5&1.5&0.0&0.5\\
    3&1.8&0.0&0.3&1.0&0.0&0.3&0.2&0.0&0.5&0.3&0.0&0.0\\
    4&0.0&0.0&0.9&0.4&-0.1&0.2&0.2&0.0&0.0&0.4&0.0&0.1\\
    5&1.4&0.2&1.2&2.3&-0.1&0.9&0.4&0.0&0.5&0.7&0.0&0.8\\
    6&20.0&1.9&49.2&16.4&-1.2&37.9&28.0&8.7&56.2&24.1&3.3&38.6\\
    7&2.4&-0.1&1.0&1.1&-0.3&0.4&0.2&0.0&0.4&0.2&0.2&0.5\\
    8&1.5&0.0&1.0&1.1&-0.1&0.8&0.3&0.4&1.4&1.5&0.2&0.5\\
    9&2.8&-0.1&0.3&-0.3&0.0&0.2&0.0&0.0&0.2&-0.6&-0.1&0.3\\
    10&2.1&0.1&3.2&2.3&0.4&1.4&0.2&-0.2&0.8&0.9&0.1&0.7\\
    11&1.2&0.1&0.8&0.6&0.0&0.3&0.2&0.1&0.5&1.4&0.0&0.5\\
    12&9.4&-0.1&3.2&4.4&-0.1&1.9&0.4&0.1&0.9&2.4&0.1&1.9\\
    13&2.3&0.1&1.9&17.9&0.8&18.1&0.3&-0.2&1.3&4.3&0.0&1.3\\
    14&17.7&0.9&24.1&16.7&-0.9&34.0&17.2&0.0&27.6&26.4&1.9&50.4\\
    15&-0.4&-0.1&0.2&0.4&0.0&0.1&0.8&0.0&0.3&0.4&0.0&0.1\\
    16&0.8&0.0&0.3&0.2&-0.1&0.2&0.0&0.0&0.8&2.4&0.1&0.5\\
  \hline
  \end{tabular}
\label{tab:tableDistrictResults}
\end{table}

Observing Table \ref{tab:tableDistrictResults} one may see that different taxi companies can have Shapley values that differ by several orders of magnitude  within the same district. Also, the Shapley value of a given company may vary from district to district by a factor of more than $\times 10$ in some cases (see, for instance, companies 1 and 13 in districts 15, 17 and 19). Some companies have negative Shapley values in certain districts, meaning that they are bringing \textit{on average} a negative contribution (\ie reduce the forecast accuracy) to the coalitions they join. 

From Fig.~\ref{Figure 5} we see that the Shapley values of companies do not correlate well with their number of rides. In fact, the Shapley value for small companies tends to be higher than their corresponding percentage of rides, whereas it is the opposite for large companies. In other words, if we approximated the importance of different companies just by the volume of data (rides) that they contribute, we would be rewarding large companies, at the expense of smaller ones.

Similarly, LOO values are weakly correlated ($R^2 = 0.38$) with the corresponding Shapley values. Notice that even some big companies (large number of reported rides) have negative LOO (e.g., company 6 in district 15). As with the case of number of rides, this means that if someone were to split an accepted bid based on LOO values, the allocation of payments to different companies would deviate significantly from what a splitting based on Shapley would produce.

\subsection{Summary}
Predicting demand at city level does not require collaboration between different taxi companies since each one can independently estimate city-wide demand. However, when attempting to estimate demand at district level, different companies need to combine their data if they are to achieve a high prediction accuracy. In these cases, neither the data volume a company is providing nor its LOO value reflect accurately its contribution to achieving a better forecast of future demand as given by its corresponding Shapley value.

\section{Computing the importance of data in retail markets}
\label{sect:ValueForIndividual}

In the previous section we developed methods for estimating the value of aggregate data held by taxi companies. In this section we will go a step further, and develop methods for estimating the value of data held by individual drivers. This will introduce additional challenges in terms of scalability of computation, since the Shapley value will now have to be computed over hundreds or thousands of individual taxi drivers. 

\subsection{Selling spatio-temporal data through a PIMS}
\label{subsec:PIMSdef}

As discussed on the introduction, in the last few years several Personal Information Management Systems (PIMS) have appeared that allow individuals to sell their data directly in a ``retail'' data marketplace. To conduct our study of estimating the value of information held by individual drivers, we will assume a simple model of such a PIMS. The design space for PIMS is of course huge, but it is beyond the point of this paper to examine the different alternatives. In Section~\ref{sec:conclusions} we discuss several important aspects for implementing real-world data marketplaces, both for retail and wholesale data. In the rest of this section we will assume that the PIMS operates as follows:

\begin{itemize}
\item Drivers upload to the PIMS their rides each day.
\item Buyers request from the PIMS to train their forecasting algorithm for spatio-temporal demand using data from real drivers.
\item The PIMS uses a sufficient number of drivers' data to reach an accuracy threshold set by the customer.  
\item Buyers pay the PIMS.
\item The PIMS keeps a small percentage of the payment and returns the remaining part to the drivers whose data was used in training the buyer's forecasting algorithm.
\end{itemize}

We will assume again that the PIMS uses the  forecasting algorithm described in Sect.~\ref{sect:problem description}. To achieve the requested accuracy, the PIMS starts with a random number of drivers, trains the algorithm, and computes its accuracy over a test set. If the accuracy threshold is not reached, then the PIMS selects an additional set of drivers until it gets to the desired accuracy, or fail to do so, in which case it informs the buyer that the request cannot be met. 

\subsection {City-wide results}
We have computed a TSS Shapley value approximation for a set of $|N| = 4968$ taxi drivers that provided service in Chicago during March and April 2019. We sampled $r=8$ different permutations for each driver and applied a $Truncation\_threshold = 0.95$. In this way we computed the contribution of each driver's data to the forecasting accuracy achieved by the multiseasonal SARIMA model in predicting the demand in the second half of April using taxi rides from the previous six weeks for training ($T_o = $ Mar. 4th - Apr. 14th and $T_c = $ Apr. 15th - 28th). 

\begin{figure}[tbp]
  \centering
  \includegraphics[width=300pt]{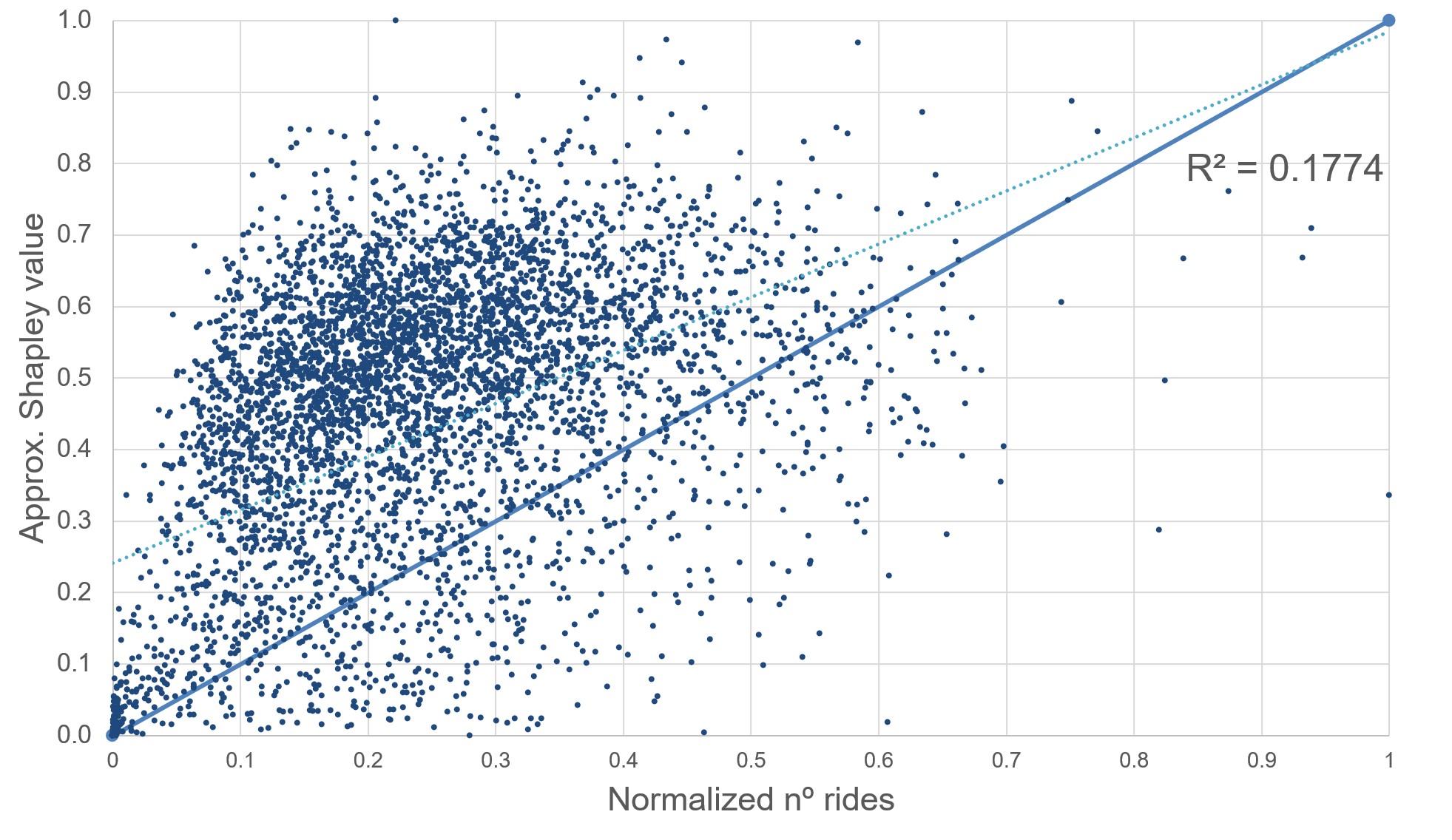}
  \caption{Approximate Shapley value vs. the number of rides across drivers}
  \label{fig:Figure6a}
\end{figure}

In the same way that we proceeded in the wholesale use case, we compared the Shapley value with the number of rides reported by each driver. Figure~\ref{fig:Figure6a} shows a plot of these two metrics across all drivers. We see that there is no clear relationship between them. In fact, the linear correlation between both values is very low $(R^2=0.1774)$. Another interesting finding is that it takes a very small number of drivers to estimate the city-wide aggregate demand. With 7 randomly selected drivers, on average, we can reconstruct the shape of the demand at city level with a 95\% accuracy. 

\subsection{District-level results}
\label{sect:DoWeNeedAllUsers?}
\begin{figure}[tbp]
  \centering
  \includegraphics[width=8cm]{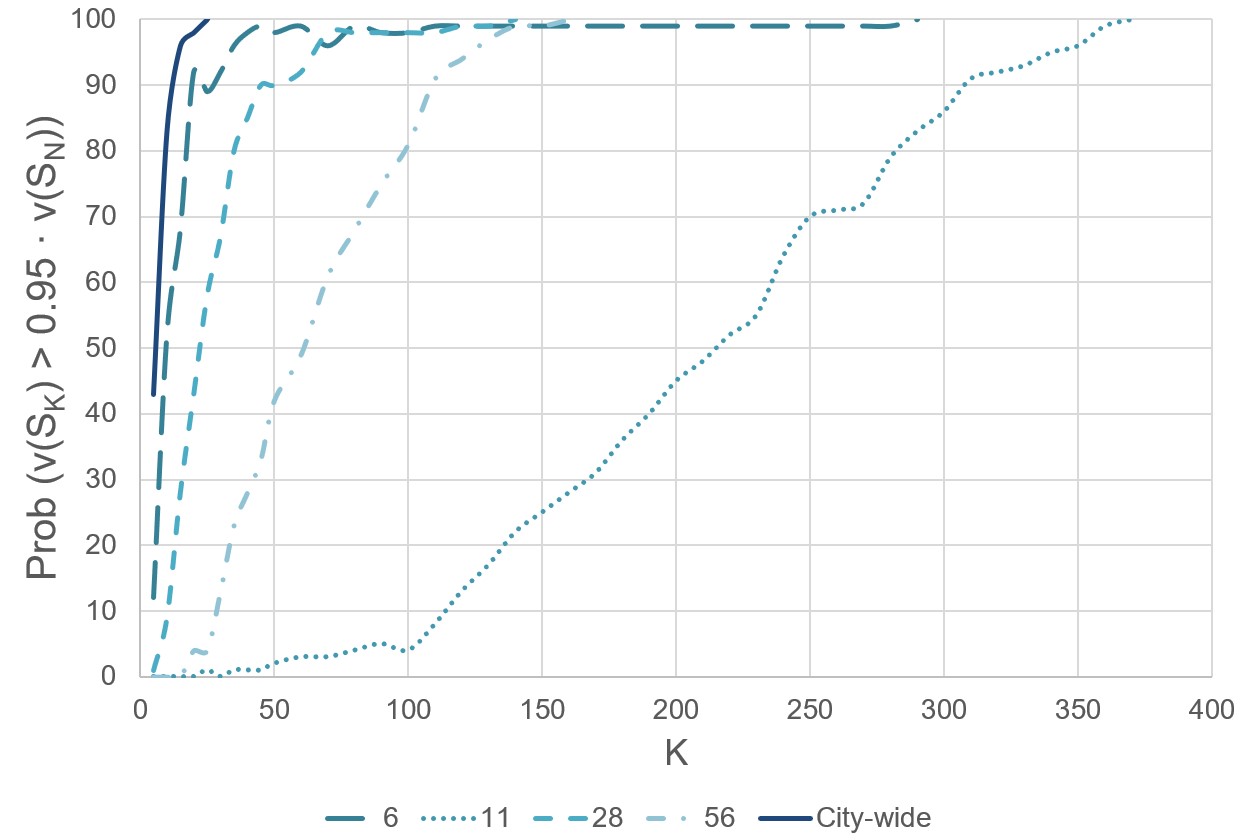}
  \caption{ Probability of $v(S_K)$ exceeding $95\% \cdot v(S_N)$ vs the number of drivers in K }
\label{Figure 7}
\end{figure}

As we just saw, it is possible to build accurate demand forecasts at city level using only a very small number of drivers. But what if a customer asks for demand forecasts at the district level? 
To address this, we will first quantify the number of necessary drivers, and then proceed to compute the relative value of each driver's data. Figure \ref{Figure 7} shows the probability that using a number of drivers indicated in the x-axis one can achieve a prediction accuracy at least 95\% of that achieved when using information from all the drivers. Different lines correspond to districts with high (28), medium (6 and 56) and small (11) demand for taxi rides.

The plot shows that whereas for forecasting city-wide demand, or demand in large districts, few drivers suffice, forecasting the demand of medium-sized and smaller districts requires information from many more drivers. This can be understood by noting that in large districts, the aggregate demand is much more predictable since it is the result of the aggregation of large numbers of independent variables (people that may need a taxi ride). Such demands are known to be easier to forecast (see for example~\cite{Laoutaris09} in which the traffic of large backbone network links is easier to predict than the traffic of smaller access links). Achieving high forecasting accuracy in medium and small districts requires using the data from tens if not hundreds of drivers. Computing the actual Shapley value is impractical for such numbers of players, but it can be approximated by using the structured sampling approach discussed earlier in Sect.~\ref{sect:computing Shapley of individuals}.

\begin{figure}[tb]
  \centering
  \includegraphics[width=8cm]{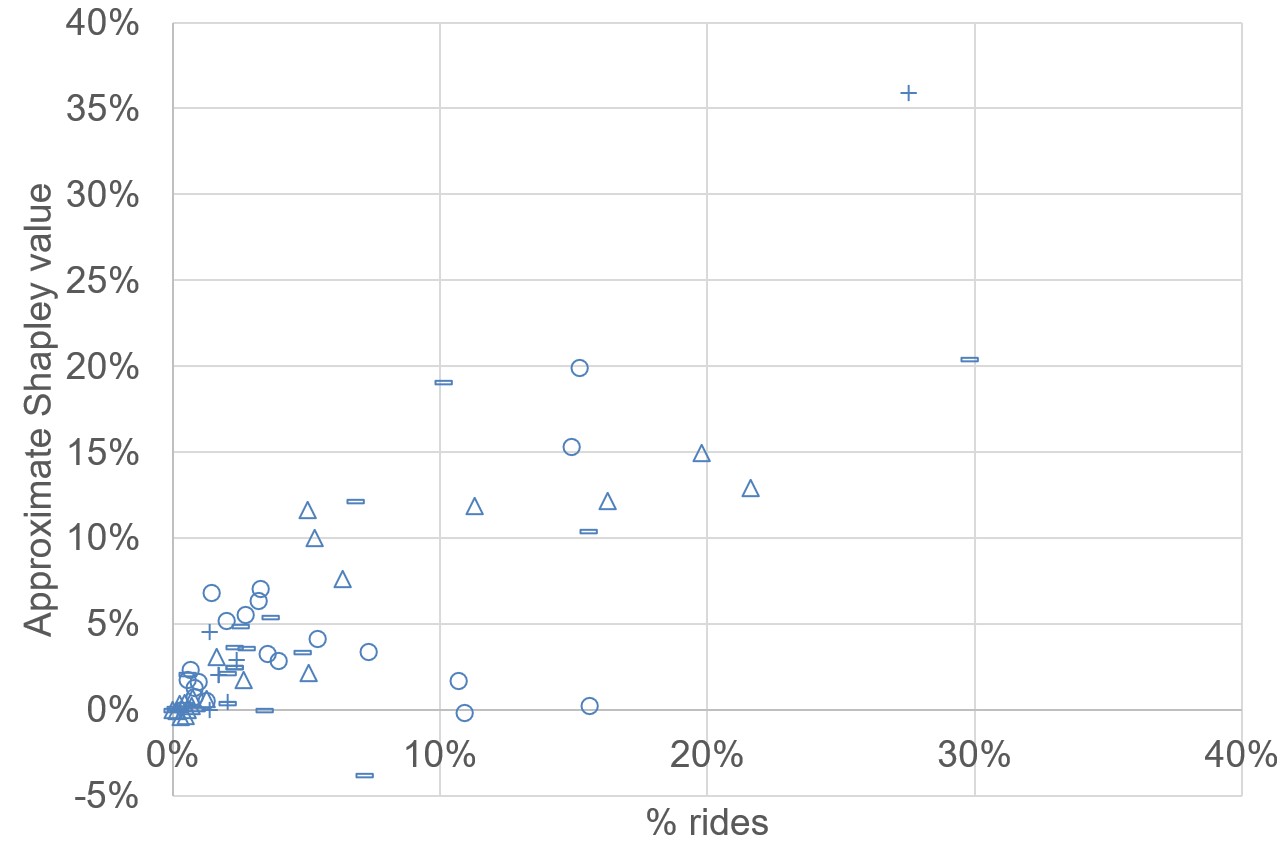}
  \caption{ Scatter plot of approximated $\phi_i$ vs nº rides (\%) for tuples of 20 drivers in district 28 which yield more than $0.95 \cdot v(S_N)$ prediction accuracy.}
\label{Figure 8}
\end{figure}

We computed the Shapley values for smaller sets of drivers whose data achieve an accuracy very close to $v(S_N)$ when combined. Figure \ref{Figure 8} shows a scatter plot of the approximate Shapley value (Y-axis) vs. percentage of reported rides (X-axis) for a number of such sets of drivers in district 28. Each point represents a driver, and drivers from the same set are represented with the same marker. As observed earlier at city-level, the real value of a driver may be very different from that predicted by its number of rides.

\section{Taxi demand forecasting in NYC}
\label{sect:TaxiNYC}
We have repeated the analysis using this time a dataset of taxi rides that took place in New York City from April to May 2019 \footnote{see \url{https://www1.nyc.gov/site/tlc/about/tlc-trip-record-data.page}, last accessed January 2020}. The dataset includes more than 65 million rides from 33 companies in 261 districts.

The conclusions from NYC are similar to the ones we drew in detail for Chicago. Particularly in the case of NYC, more than 80\% of taxi companies are able to predict demand with an accuracy of above 80\% in 229 districts. Cooperation can improve by at least 10\% the accuracy of individual forecasting for more than 75\% of the companies in 27 of the smallest districts. There are 4 districts with very few rides in which the forecasting algorithm cannot achieve a high accuracy even if it combines the data from all companies. In the districts where cooperation between companies made sense, the number of rides reported by each company is again weakly correlated with its Shapley values, as was also the case in Chicago. 

Similar conclusions are obtained when analysing the value of individual drivers. Their Shapley value does not correlate well with the number of rides reported by each driver ($R^2$ ranging from 17\% to 40\%) and the same holds for their LOO value. In conclusion, repeating the analysis for a second large dataset verified all our main conclusions obtained from the analysis based on the Chicago dataset. 

\section {Related works}
\label {sect:RelatedWorks}

The use of spatio-temporal data in transportation and smart city applications has attracted much attention from the research community. Works like~\cite{Xia19, Zhao19, Fang18} look at how knowledge extraction from spatio-temporal data can improve the effectiveness of transportation \cite{Yan18} and delivery services \cite{Zhang19}. The research community has developed several data fusion methods for combining data from different domains with different structure. \cite{Zheng15} provides a comprehensive summary on data fusion, including examples of fusing spatio-temporal data with map data for improving road networks \cite{Zheng11, Pan13} or diagnosing traffic anomalies \cite{Liu11, Chawla12}.

Despite the large literature on data fusion, we are aware of only a couple of papers that study data valuation matters around spatio-temporal data~\cite{Aly18, Aly19}. These works differ from ours in at least the following aspects: 1) they employ empirical notions of value instead of the Shapley value used in our work, 2) they are concerned with variations on the value of data with time, and with identifying the best time for making a bid, and 3) they look at different applications domains (e.g, location-based marketing) than the one that we study.  

In addition to the commercial marketplaces discussed on the introduction, the research community is also actively working on proposing novel data marketplace architectures ~\cite{Agarwal19, Chen19, Babioff12} and pricing policies for data~\cite{Mehta19}. Most of these works are theoretical in nature, focus on bidding strategies for buyers, and do not answer any of the main questions addressed in our work.

Another body of related work has to do with computational aspects of the Shapley value across different application domains. \cite{Ghorbani19} has used the Shapley value to compute payments for providers of training data for different machine learning problems (not related to spatio-temporal demand prediction). Several works have looked at computational aspects of Shapley value and for efficient exact and approximation algorithms for particular types of problems such as recommendation, graph centrality, and others~\cite{Jia19, Ghorbani19, Cabello18, Paraschiv19}.

The idea of providing \textit{micropayments} to users for their personal data has received a lot of public attention after the publication of "Who owns the future?" by Jaron Lanier in 2013~\cite{Lanier13}. More recent work describes fundamental technological challenges that need to be addressed for the above vision to be fulfilled~\cite{Laoutaris19}. None of the above works has looked at valuation issues relating to spatio-temporal data.

\section{Conclusions and future work}
\label{sec:conclusions}
In this work we have looked at the problem of how to compute the relative importance of different spatio-temporal datasets that are combined in order to improve the  accuracy of demand forecasting for taxi rides in large metropolitan areas such as Chicago and New York. Our main result has been that the importance of each dataset cannot be deduced via simple heuristics based on data volume or leave-one-out methods, but instead one needs to look deeper and consider the complex ways in which different datasets complement one another, which is what the Shapley value does. This applies when combining data from entire companies, as well as when combining data from individual drivers.

In this work we have addressed our main question in a ``full information'' setting, i.e., assuming that we can test the forecasting potential of each individual dataset or coalition of datasets. This was done in order to answer our main question regarding whether simple or more complex methods needs to be used in order to split the value of a dataset in a fair manner. Designing a fully functioning marketplace that implements our ideas is a much bigger problem that we are currently working on. Several additional challenges remain in order to achieve this. For example:

\begin{itemize}
    \item Data buyers need to have a way to estimate the value of a coalition of datasets for their machine learning algorithm, without, however, having access to raw data that they have not purchased yet. This can be done in different ways, including having the marketplace execute the training locally and communicating to data buyers the achieved accuracy.
    \item Data buyers also need to be protected against strategic data sellers that may modify their datasets artificially in order to receive higher bids.
    \item Once a bid has been accepted by the sellers, splitting it in proportion to the different Shapley values is a reasonable and fair approach but need not be the only one. To attract new sellers, and to retain existing ones, a marketplace may choose to pay a minimum amount to some sources, even if their Shapley values do not justify this. Other considerations may also prompt it to partially deviate from splitting according to Shapley value.  
    \item Partial information models in which not even the marketplace has full information on each and every dataset can also be considered. 
\end{itemize}

In addition to the above, we are also looking at value sharing in the context of more complex metrics than just the spatio-temporal footprint of served demand. Such metrics include full origin-destination traffic matrices as well as congestion in road networks which will require data not only from different sources, but also from different domains. Last, we are working on developing approximation algorithms for computing the Shapley value in spatio-temporal and other settings. 

% ---- END OF PAPER ----

% Appendices
\appendix

\section{Appendix}

\subsection{Testing Shapley value approximation algorithms}
\label{sect:ShapleyApproxAlgorithms}
The aim of this appendix is to introduce the algorithms that were analyzed and evaluated to select the most suitable approximation to the Shapley value. Since an exact calculation of the Shapley value requires an $O(2^{|N|})$ algorithm, the performance of the approximation algorithms to be used is critical. Having evaluated different candidate algorithms extensively, we have selected a structured sampling algorithm for it provides the best trade-off between accuracy and time on all the datasets we have tested. Even though it is tailored to the behaviour of the value function defined for this specific problem, it will outperform naive methods in any problem where the marginal contribution of a player to a coalition strongly depends on the number of players in such coalition.

\subsubsection{Explaining the evaluation testbed}

We computed the exact Shapley value for the daily prediction model, using a wholesale setting as described in section \ref{sect:wholesale setting}, in order to test the accuracy of the prediction algorithms, both at city level and in a medium-size district (particularly district 35). The following approximation algorithms were evaluated:
\begin{enumerate}
    \item Monte Carlo (hereinafter, MC) approximation as stated in \cite{Ghorbani19} evaluates the marginal contribution to coalitions extracted from a random sample of permutations of $N$ until a convergence condition is met. We selected as such convergence condition a flag that controls whether the maximum relative variation of approximated $\phi_i$ is below an input threshold, which will range from 10\% to 0.5\%, before computing a new permutation. 
    \item Random Sampling (hereinafter, RS) as stated in \cite{Castro09} using a different number of $r \cdot |N|$ sample permutations, where $r$ will range from $r = 1$ to $r = |N|^2$.
    \item Structured Sampling (hereinafter, SS), tailored to problems where the position of \textit{players} in a permutation strongly determines their marginal contribution, based on \cite{Fatima08, vanCampen17}. SS ensures that all companies appear $r$ times in each position for a set of $r \cdot |N|$ permutations.
\end{enumerate}

Given the stochastic nature of Shapley approximation algorithms, we tested each one 50 times for each set of input parameters and obtained the approximate Shapley value for the 16 companies. We compared the performance of the aforementioned algorithms in terms of:
\begin{itemize}
    \item Accuracy, measured as the average average\footnote{First average error across companies for each test, then average the average error across all executions} absolute error (hereinafter, AAAE) and average average percentage error (hereinafter, AAPE) compared to the exact Shapley Value by company.
    \item Robustness, measured as the average average\footnote{First average standard deviation of the approximate Shapley value across all executions, then average the average standard deviation across companies} standard deviation (hereinafter, AASTDE) of the outputs of an algorithm and a certain set of parameters.
    \item Time to execute (TtE), measured in terms of the number of training-prediction cycles computed.
\end{itemize}

The convergence threshold in the case of MC and $r$ in the case of RS and SS allow one to define the sample depth and affect both to the execution time and to the accuracy of the approximation.

In all cases, we tested non-truncated and truncated versions of the MC, RS and SS algorithms (we will refer to them as truncated-XXX algorithms, or in short-form TMC, TRS and TSS). Truncation of execution above a certain \textit{truncation threshold} ($v(S_N) - \epsilon$) works in the following way: while evaluating a permutation $\pi$ of the set $N$, if it holds that for the coalition of the first j players, $\pi_j  \subseteq \pi, \pi_j = \{\pi[1], ... , \pi[j]\}, v(S_{\pi_j}) > v(S_N) - \epsilon$, the rest of members $k \in \pi - \pi_j$ are considered to bring a zero marginal contribution. Truncation helps the algorithm reduce the time it takes to execute but also decreases the accuracy of the approximation.

\subsubsection{Explaining structured sampling (SS) algorithm}
\label{Explain ISS}

As regards \textit{"structured sampling"} (SS) approximation to Shapley value, the \textbf{Algorithm 1} box provides a detailed description of the algorithm. Unlike random sampling, it plans the sample permutations upfront so as to ensure that each player $i \in N$ appears $r$ times in each position of the sampled permutations. This reduces the randomness of the sampling process and increases the performance, especially when the marginal contribution of a player i in a permutation $\pi$ is significantly determined by its position.

\begin{algorithm}
\caption{Structured sampling approximation algorithm}
\begin{algorithmic}[1]
    \STATE \textbf{inputs}: $S_{n_i}$ train data for each $n_i$ source in the set N, accuracy test procedure $v$, rounds of permutations to evaluate $r$ and truncation threshold $\epsilon$
    \STATE Initialize Shapley value vector $\phi_i = 0, \forall i \in N$
    \STATE Initialize the set of sample permutations $P = \emptyset$
    \STATE Create a $|N| x |N|$ Latin square - $LS$
    \STATE $Q \leftarrow N$
    \FORALL{$i \in \{1...r\}$}
        \STATE $Q \leftarrow \text{shuffle} (Q)$
        \STATE {$P \leftarrow$ set of $|N|$ permutations of $Q$ according to the order defined by $LS$}
    \ENDFOR
    \STATE $t \leftarrow 0$
    \FORALL {$\pi^t \in P$}
        \STATE $t \leftarrow t + 1$
        \STATE $v_{j-1} \leftarrow 0$
        \FORALL {$j \in \{1...|N|\}$}
            \IF {$v^t_{j-1} <= v(S_N) - \epsilon$}
                \STATE $v^t_j \leftarrow v(S_{\pi^t_j})$
            \ELSE
                \STATE $v^t_j \leftarrow v^t_{j-1}$
            \ENDIF
            \STATE $\phi^t_{\pi^t[j]} \leftarrow \frac{t-1}{t} \cdot \phi^{t-1}_{\pi^t[j]}+\frac{1}{t} \cdot (v^t_j - v^t_{j-1})$
        \ENDFOR
    \ENDFOR
    \STATE \textbf{outputs}: Approximation to the Shapley value of each data source i: $\phi_1 ... \phi_{|N|}$
\end{algorithmic}
\label{algo2}
\end{algorithm}

We resort to Latin squares in the sample process. A Latin square $LS$ of order $|N|$ is an $|N| \times |N|$ array with elements of a set $N$, in such a way that each element $n_i$ occurs precisely once in each row and column of the array \cite{Jacobson96}. Latin squares have been extensively used in experiment planning \cite{Williams49}. As shown in the \textbf{Algorithm 1} box, we shuffle the elements of a random permutation $Q$ of $N$ according to the order defined by such Latin square to produce $r$ different sets of $|N|$ permutations with the aforementioned properties.

The referenced shuffle algorithm is the modern version of the Fisher–Yates shuffle, designed for computer use by Richard Durstenfeld \cite{Durstenfeld64}. Such algorithm runs in $O(n)$ time and is proven to be a perfect shuffle, assuming a reasonably good random number generator.

\subsubsection{Evaluating Shapley value approximation algorithms}
\label{sect:EvaluatingApproxAlgorithms}
\begin{figure}[tb]
  \centering
  \includegraphics[width=\linewidth]{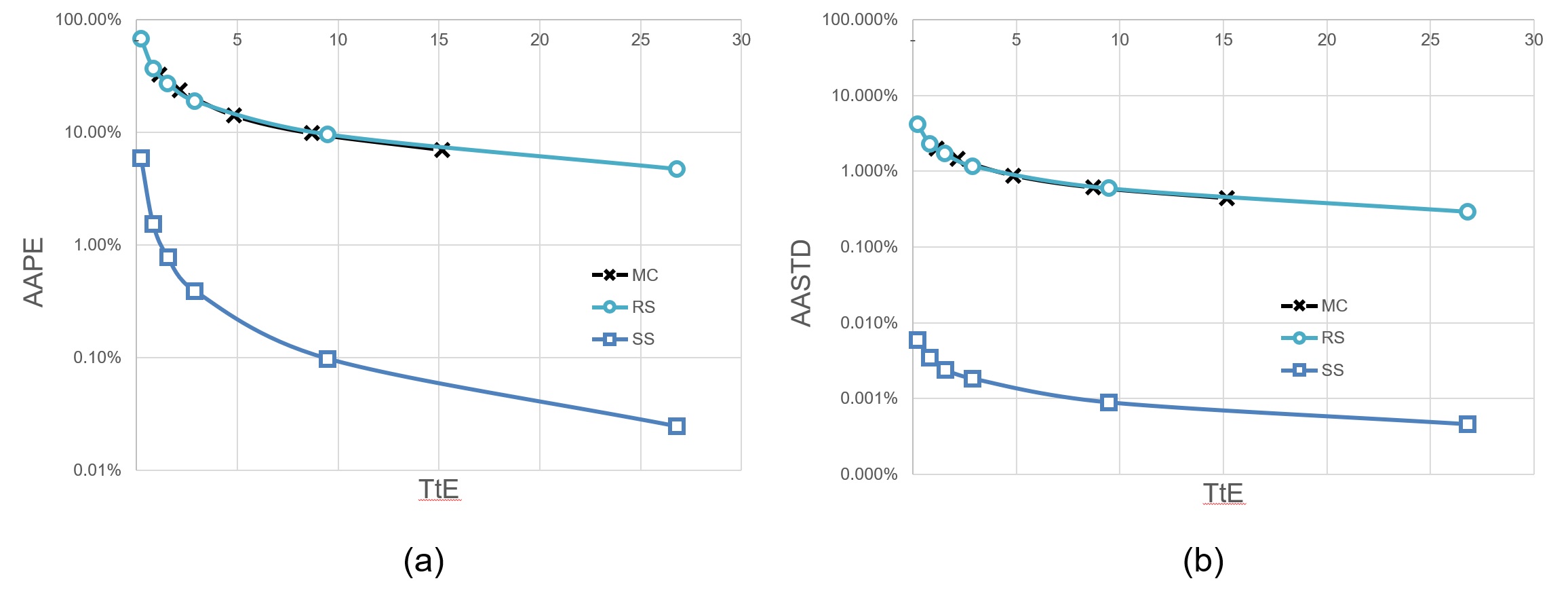}
  \caption{ Testing of Shapley value approximation algorithms for demand prediction models at city level (|N|=16). (a) Accuracy (AAPE) \vs execution time. (b) robustness \vs execution time}
\label{FigureSVAlgorithmCity}
\end{figure}

Fig.~\ref{FigureSVAlgorithmCity} shows a comparison of MC, RS and IS in terms of accuracy and robustness. In subplot (a) we depict the AAPE as a function of the number of sub-coalitions evaluated, which determines execution time. Subplot (b) shows the AASTD as a function of the execution time. Please note that Y-axis is logarithmic in both subplots.

In all cases the more combinations are evaluated, the more accurate and, especially, more robust the results are. Nonetheless, it is clearly shown in this case that the SS outperforms both RS and MC, meaning that the planning of the sample permutations delivers a consistent output across executions which is also closer to the exact Shapley values.

\begin{figure}[tb]
  \centering
  \includegraphics[width=\linewidth]{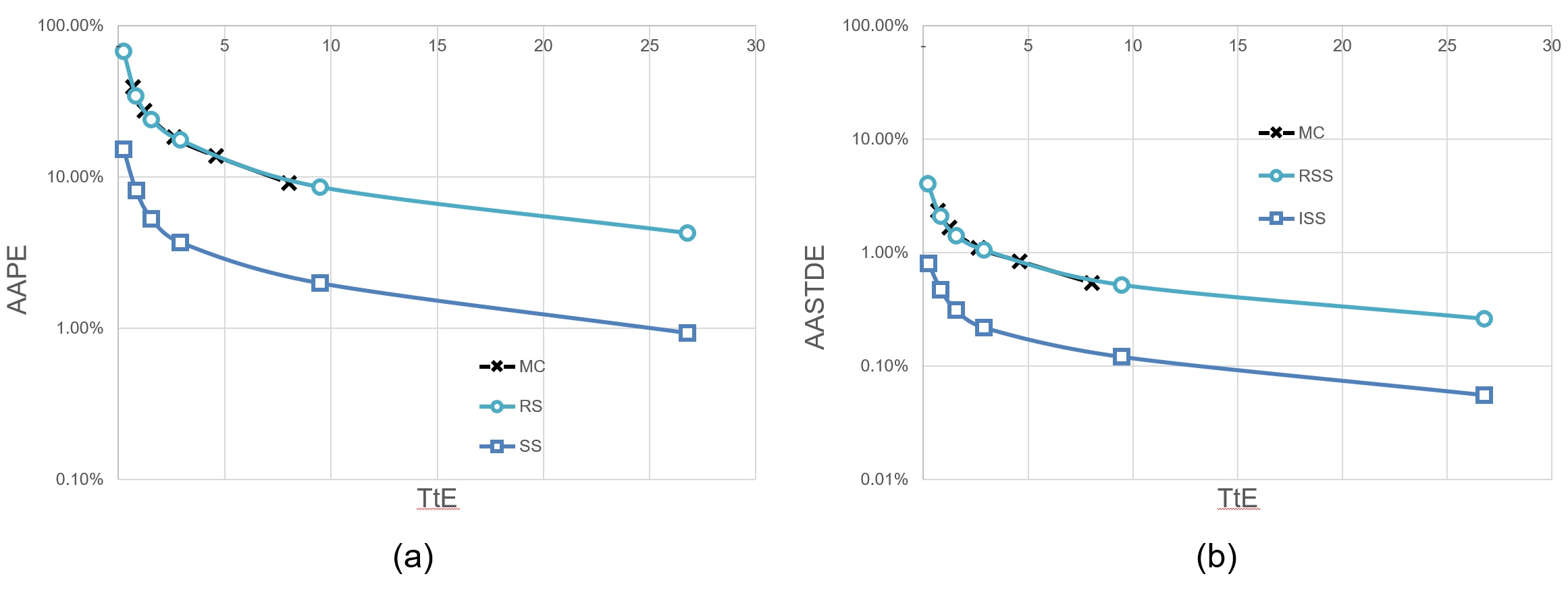}
  \caption{ Shapley value approximation algorithm testing for demand prediction models for district 35 in Chicago and N=16 companies. (a) Accuracy (AAPE) \vs execution time. (b) robustness \vs execution time.}
\label{FigureSVAlgorithmDistrict}
\end{figure}

Since city wide demand prediction considering data from companies is a very special case \footnote{If we recall section \ref{sect:wholesale city}, all companies have enough data to independently predict the shape and average mean of the aggregate demand, with at most a 10\% maximum average error, meaning that the company that appears in the first place in the permutation is bringing all the value. This might be a best case for SS.}, we ran the same test using the inputs of a medium-size district, which is yielding very different Shapley values for each company, to prove whether or not the previous conclusions hold, in a scenario where the standard deviation of $\phi_i$ across companies is relevant. Figure \ref{FigureSVAlgorithmDistrict} shows the results of this analysis.

As expected, the difference between SS and the naive versions of MC and RS in terms of robustness and accuracy decreases in cases where $\phi_i$ values are very different. MC takes more time to converge and both RS and SS results show higher AAPE if compared with the first most favourable case. However, the SS algorithm showed to clearly outperform both MC and RS also in this situation.

In the light of the results obtained, we have selected SS as the best algorithm, since it is able to to approximate the Shapley value with a 10\% average error in $O(|N|^2)$. This we consider sufficient for the purpose of a value-based payoff distribution. In case finer accuracy is required, SS is able to estimate the Shapley value with a 4\% of error in $O(|N|^3)$.

\subsubsection{Evaluating the impact of truncation on accuracy}

For computing the Shapley value for a large number |N| of \textit{players}, and given the specific behaviour of value in demand prediction problems, truncation proves to be an important feature to speed up the execution without necessarily distorting the output of the algorithm. In fact, if according to section \ref{sect:ValueForIndividual} by taking only a small percentage of all drivers we are able to quite accurately predict demand in most cases, then why spending our valuable computing time evaluating the marginal contributions of additional \textit{players} once our prediction has reached a $95\%$ of the maximum accuracy?

We have computed approximations to Shapley value for the following truncation thresholds: 0.6, 0.7, 0.8, 0.9, 0.95, 0.97, 0.98, 0.99 and 1. We have run TMC, TRS and TSS in the same district as we did in section \ref{sect:EvaluatingApproxAlgorithms} for $|N| = 16$ companies, 50 times for each $\epsilon$ and using  a convergence threshold of $0.01 \cdot v(S_N)$ for TMC and  $r = 64$ for TRS and TSS.

\begin{figure}[tb]
  \centering
  \includegraphics[width=\linewidth]{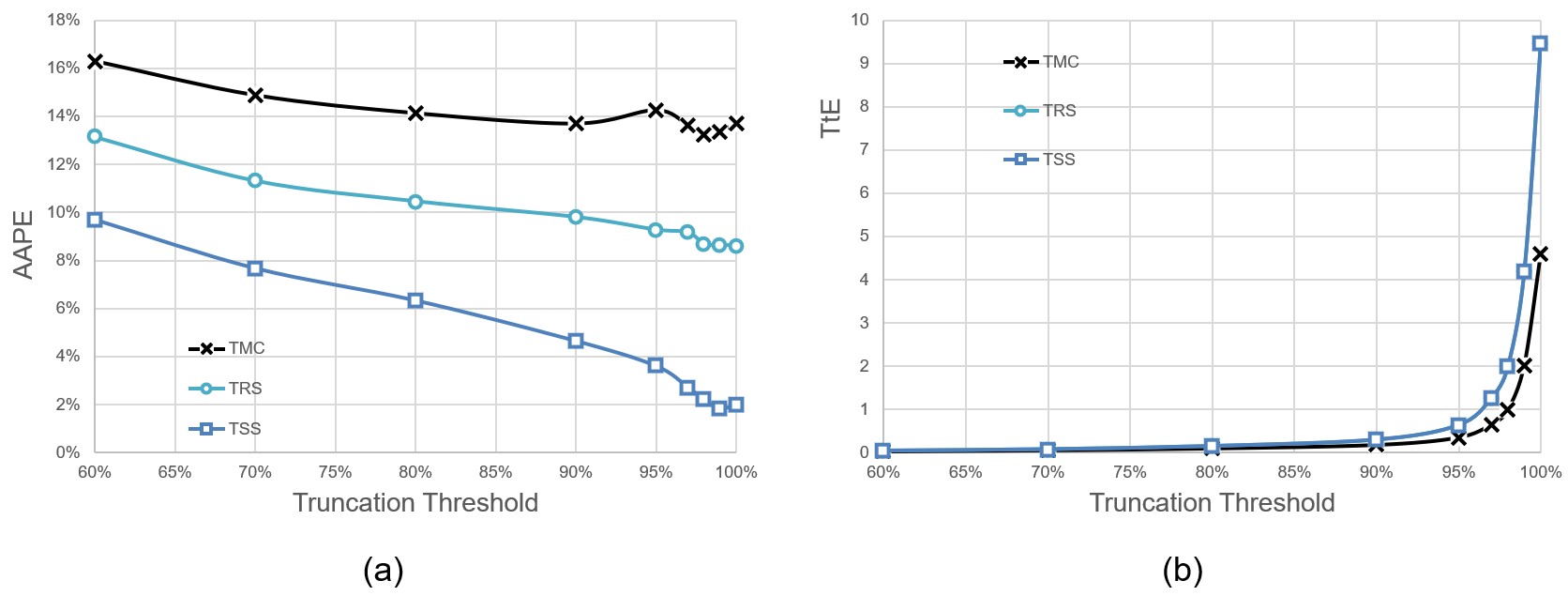}
  \caption{ accuracy and TtE of $\phi_i$ vs. truncation threshold in approximation algorithms. In this case, SS and RS overlap in the (b) plot. }
\label{FigureSVAlgorithmTruncation}
\end{figure}

Figure \ref{FigureSVAlgorithmTruncation} shows the effect of truncation on both accuracy (a) and execution time (b) for each of the three algorithms. According to our results, SS is significantly more sensitive to truncation, but it is possible to easily control the trade-off between accuracy and execution time by tuning $r$ and $\epsilon$. We chose to use a \textit{truncation threshold} of $0.95 \cdot v(S_N)$ since it divides the overall execution time by 16, while it only duplicates the percentage error.

\subsection{Validating the results using alternative metrics}
\label{sect:validatingOthermetrics}
We have used cosine similarity (hereinafter, CosSim) throughout the paper to test the accuracy of the prediction model. As stated in section~\ref{sect:Model}, the general model is able to easily accommodate other value functions, as long as they are similarity measures, meaning by that functions $v: \hat{S_K}[t], S_N[t] \rightarrow [0,1]$, where 0 means no similarity and 1 means complete similarity between the predicted ($\hat{S_K}$) and the actual ($S_N$) served demand in the control period $T_c$. 

To validate this, we have tested the model and calculated Shapley values ($\phi_i$) using as value functions numerical similarity (hereinafter, NumSim, defined by eq.~\ref{eq:NumSim}) and relative dynamic time warping (hereinafter, RDTW, defined by eq.~\ref{eq:RDTW}) for the use case of $|N| = 16$ companies in a sample of districts with high (8, 28), medium (6 and 56) and small (11) demand. We show the results for district 11 in table~\ref{tab:tableValuationMetrics}, compared with $\phi_i$ calculated by using CosSim as the value function. We deliberately chose district 11 since it shows the highest dispersion of $\phi_i$. As it can be seen, the values using different metrics are highly correlated, and the top 4 companies are the same in the three cases. In fact, it turns out that the fraction ($\frac{\phi_i}{\sum_{j \in N}{\phi_j}}$) is very similar for the three value functions ($R^2 = 0.92$ in the case of CosSim \vs NumSim, $R^2 = 0.87$ in the case of CosSim \vs RDTW, for the 5 districts). 

\begin{table}
  \caption{Shapley value calculated using different value functions $v$ (CosSim, NumSim and RDTW) for $|N| = 16$ companies in district 11}
  \begin{tabular}{l||ccc|l||ccc}
    \hline
    Co&CosSim&NumSim&RDTW&Co&CosSim&NumSim&RDTW\\
    \hline
    1&0.02&0.01&0.02&10&0.11&0.08&0.06\\
    2&0.03&0.03&0.02&11&0.03&0.02&0.02\\
    3&0.02&0.02&0.00&12&0.10&0.07&0.08\\
    4&0.01&0.01&0.01&13&0.06&0.04&0.02\\
    5&0.04&0.03&0.01&14&0.12&0.11&0.10\\
    6&0.12&0.11&0.09&15&0.00&0.00&0.00\\
    7&0.08&0.07&0.06&16&0.06&0.04&0.03\\
    8&0.03&0.03&0.03&All&0.86&0.69&0.56\\
    9&0.01&0.01&0.01&&&&\\
  \hline
\end{tabular}
\label{tab:tableValuationMetrics}
\end{table}
\end{document}